\pdfoutput=1
\documentclass[fleqn,11pt]{wlscirep}
\usepackage[utf8]{inputenc}
\usepackage[T1]{fontenc}
\usepackage{amsmath}
\usepackage{xcolor}
\usepackage{soul}
\newcommand{\be}{\begin{equation}}
\newcommand{\ee}{\end{equation}}
\newcommand{\bea}{\begin{eqnarray}}
\newcommand{\eea}{\end{eqnarray}}

\newcommand{\p}[1]{$^2{\rm S}_{1/2} - ^2{\rm P}_{{#1}/2}$}

\graphicspath{{figures/}}
\title{Laser spectroscopy of the \p1, $^2{\rm P}_{3/2}$ transitions in stored and cooled relativistic C$^{3+}$ ions}
\author[1]{D.~Winzen}
\author[1,*]{V.~Hannen}
\author[2]{M.~Bussmann}
\author[1]{A.~Bu\ss}
\author[1]{C.~Egelkamp}
\author[3]{L.~Eidam}
\author[4]{Z.~Huang}
\author[5]{D.~Kiefer}
\author[5,6]{S.~Klammes}
\author[6,7,8]{Th.~K\"uhl}
\author[2]{M.~Loeser}
\author[4]{X.~Ma}
\author[9]{W.~N\"ortersh\"auser}
\author[1]{H.-W.~Ortjohann}
\author[6]{R.~S\'{a}nchez}
\author[2]{M.~Siebold}
\author[6,7,10]{Th.~St\"ohlker}
\author[7,9,10]{J.~Ullmann}
\author[1]{J.~Vollbrecht}
\author[5]{Th.~Walther}
\author[4]{H.~Wang}
\author[1]{Ch.~Weinheimer}
\author[6]{D.F.A.~Winters}
\affil[1]{University of M\"unster, Institute for Nuclear Physics, 48149 M\"unster, Germany}
\affil[2]{Helmholtz Center Dresden-Rossendorf, 01328 Dresden, Germany}
\affil[3]{Technical University of Darmstadt, Institute for Accelerator Science and Electromagnetic Fields, 64289 Darmstadt, Germany}
\affil[4]{Chinese Academy of Sciences, Institute of Modern Physics, 730000 Lanzhou, China}
\affil[5]{Technical University of Darmstadt, Institute for Applied Physics, 64289 Darmstadt, Germany}
\affil[6]{GSI Helmholtz Center for Heavy Ion Research, 64291 Darmstadt, Germany}
\affil[7]{Helmholtz Institute Jena, 07743 Jena, Germany}
\affil[8]{University of Mainz, Institute of Physics, 55099 Mainz, Germany}
\affil[9]{Technical University of Darmstadt, Institute for Nuclear Physics, 64289 Darmstadt, Germany}
\affil[10]{University of Jena, Institute for Optics and Quantum Electronics, 07743 Jena, Germany}
\affil[*]{e-mail of corresponding author: hannen@uni-muenster.de}
\begin{abstract}
The \p1 and \p3 transitions in Li-like carbon ions stored and cooled at a velocity of $\beta \approx 0.47$ in the Experimental Storage Ring (ESR) at the GSI Helmholtz Centre in Darmstadt have been investigated in a laser spectroscopy experiment.
Resonance wavelengths were obtained using a new continuous-wave UV laser system and a novel extreme UV (XUV) detection system to detect forward emitted fluorescence photons. 
The results obtained for the two transitions are compared to existing experimental and theoretical data.
A discrepancy found in an earlier laser spectroscopy measurement at the ESR with results from plasma spectroscopy and interferometry has been resolved and agreement between experiment and theory is confirmed. 
\end{abstract}
\begin{document}
\flushbottom
\maketitle
\thispagestyle{empty}
\section{Introduction}
The possibilities for laser spectroscopy experiments at the upcoming FAIR facility require the development of experimental tools and techniques adapted to the relativistic ion energies offered by the new accelerators. In this regard a commissioning beam-time with Li-like carbon ions at the Experimental Storage Ring (ESR) at the GSI Helmholtz Centre in Darmstadt provided the possibility to test new laser systems and a novel detection system for extreme UV (XUV) photons.
The laser systems will be required for laser cooling of bunched relativistic ion beams, as it is being planned for the SIS100 synchrotron at FAIR~\cite{Win15,Eid18}. For that purpose, continuous-wave (cw) and pulsed laser systems are developed by groups at the TU Darmstadt~\cite{Bec16} and the HZDR / TU Dresden~\cite{Sie16}, respectively. 
For the detection of forward-emitted XUV fluorescence photons that are created in the de-excitation of highly charged ions at relativistic velocities, a detector system adapted to the large Lorentz boost and Doppler shift of these photons was developed in M\"unster~\cite{Ege15}. The design is optimized for wavelengths around 10~nm as emitted in a proposed ESR experiment studying the decay of the $^3{\rm P}_1$ state in Be-like krypton ions~\cite{Win11} but is sensitive over a wide wavelength region spanning from the UV to soft x-rays.\\
In the measurements described in this paper the \p1 and \p3 transitions in C$^{3+}$-ions are investigated which are, besides their relevance for validation of theoretical atomic structure calculations (see e.g.~\cite{Ind19} or~\cite{Yer17}), of high importance for the astrophysical community where they are used for diagnostics of hot plasmas in the interstellar medium and stellar atmospheres~\cite{Gri00}.
The wavelengths of the two transitions were already determined in a previous laser-cooling experiment at the ESR~\cite{Sch05}, but were in conflict with earlier results from plasma spectroscopy~\cite{Boc63} and from interferometric measurements~\cite{Gri00}.
In this new measurement special emphasis is therefore placed on a careful calibration of all experimental components and a thorough analysis of the systematic uncertainties.
\section{Experimental setup}
\begin{figure}[h]
\centering
\includegraphics[width=0.8\textwidth]{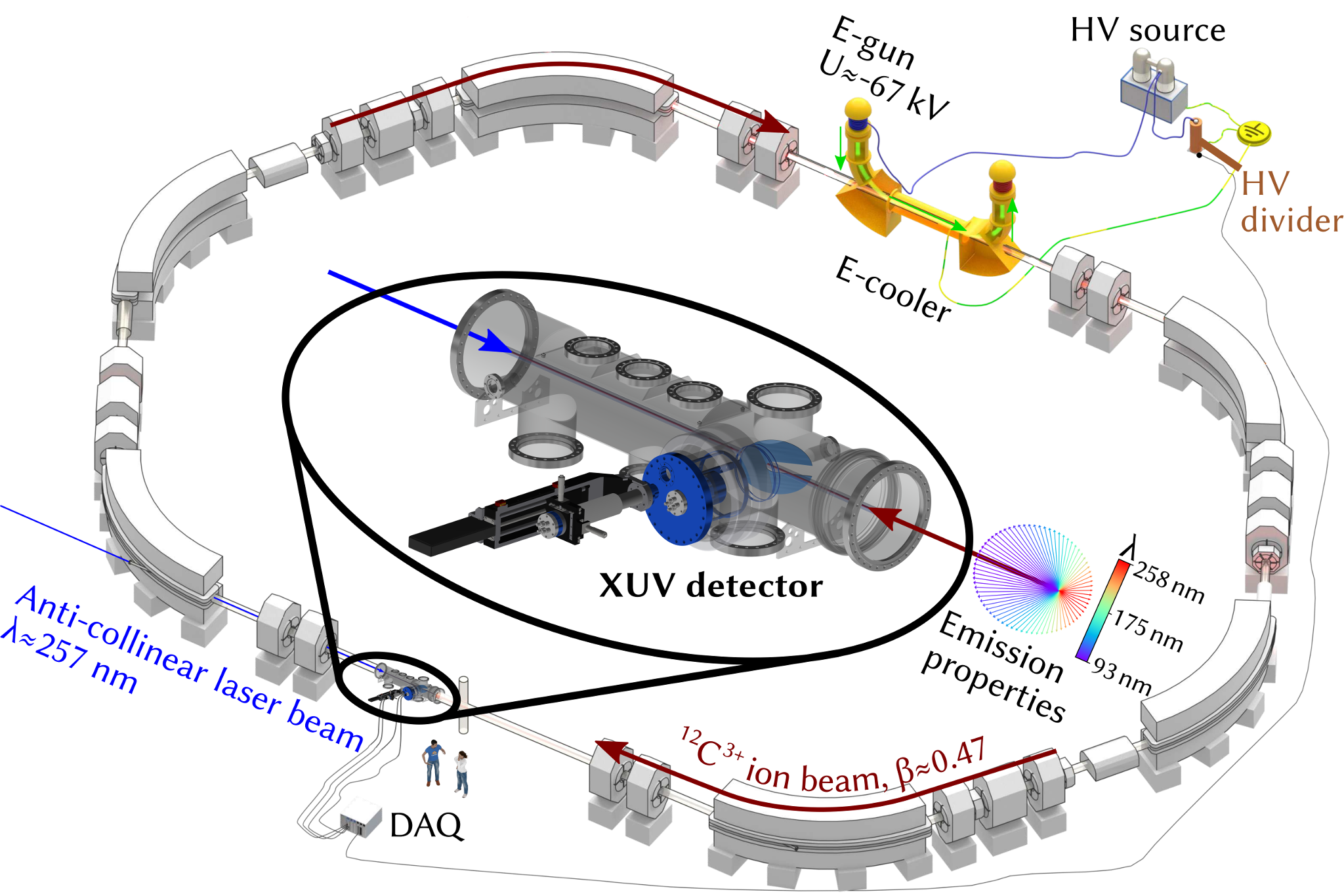}
\caption{Schematic view of the ESR with an enlarged view of the optical detection region. The ion beam is circling clockwise in the ring while the laser beam is counterpropagating with the ions inside the straight section of the ESR (figure adapted from~\cite{Ull17a}).}
\label{fig:esr}
\end{figure}

\begin{figure}[h]
\centering
\includegraphics[width=0.7\textwidth]{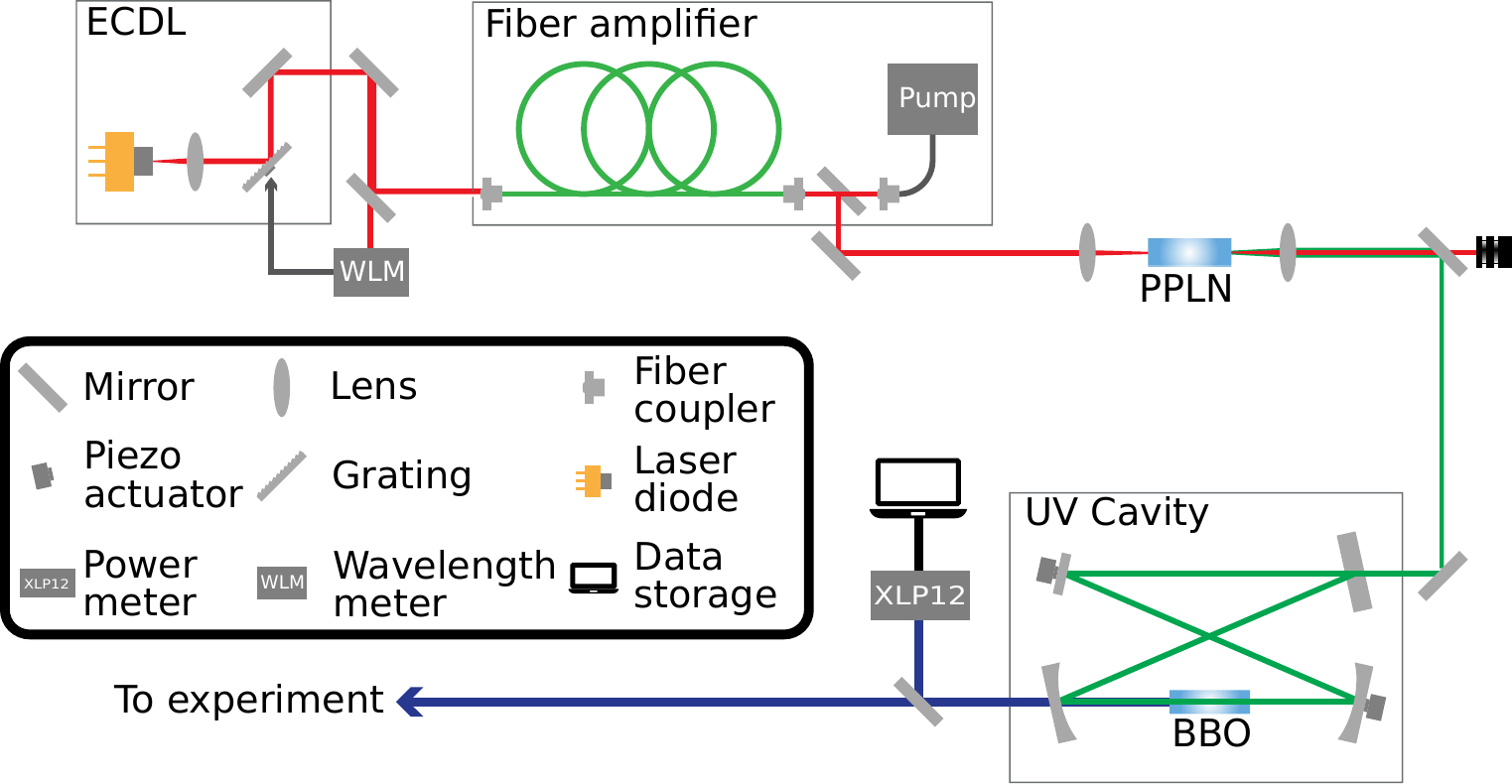}
\caption{Schematic view of the continuous-wave laser system.}
\label{fig:laser}
\end{figure}

\begin{figure}[h]
\centering
\includegraphics[width=\textwidth]{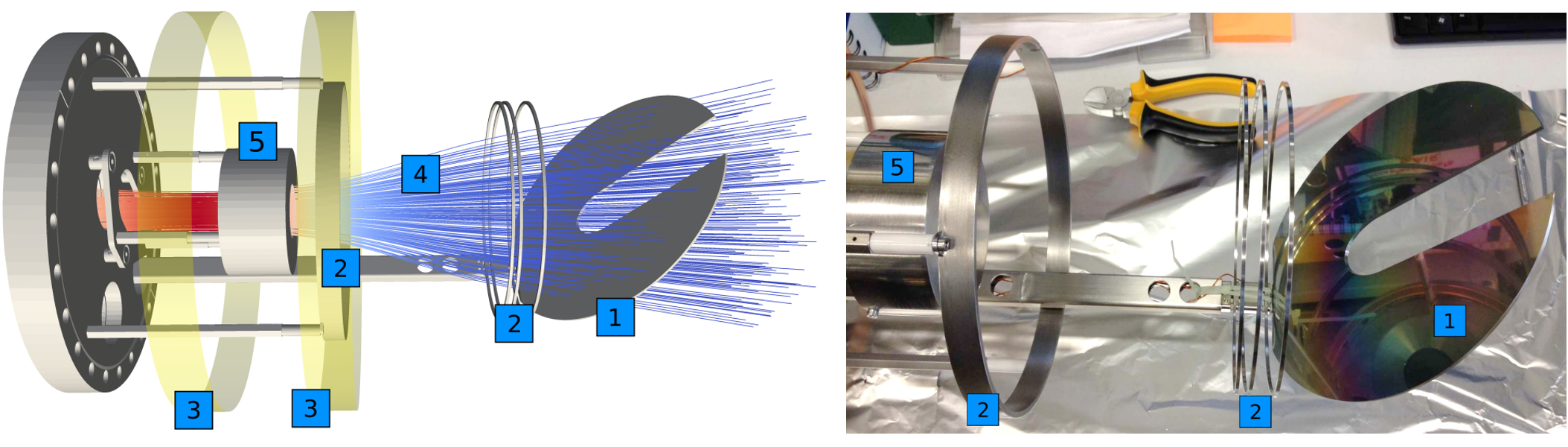}
\caption{Left: schematic view of the XUV detection system with movable cathode (1), electrodes (2), magnet coils (3), magnetic field lines (4) and MCP (5). Right: cathode plate with 300~nm CsI coating.}
\label{fig:xuv1}
\end{figure}

\begin{figure}[h]
\centering
\includegraphics[width=0.7\textwidth]{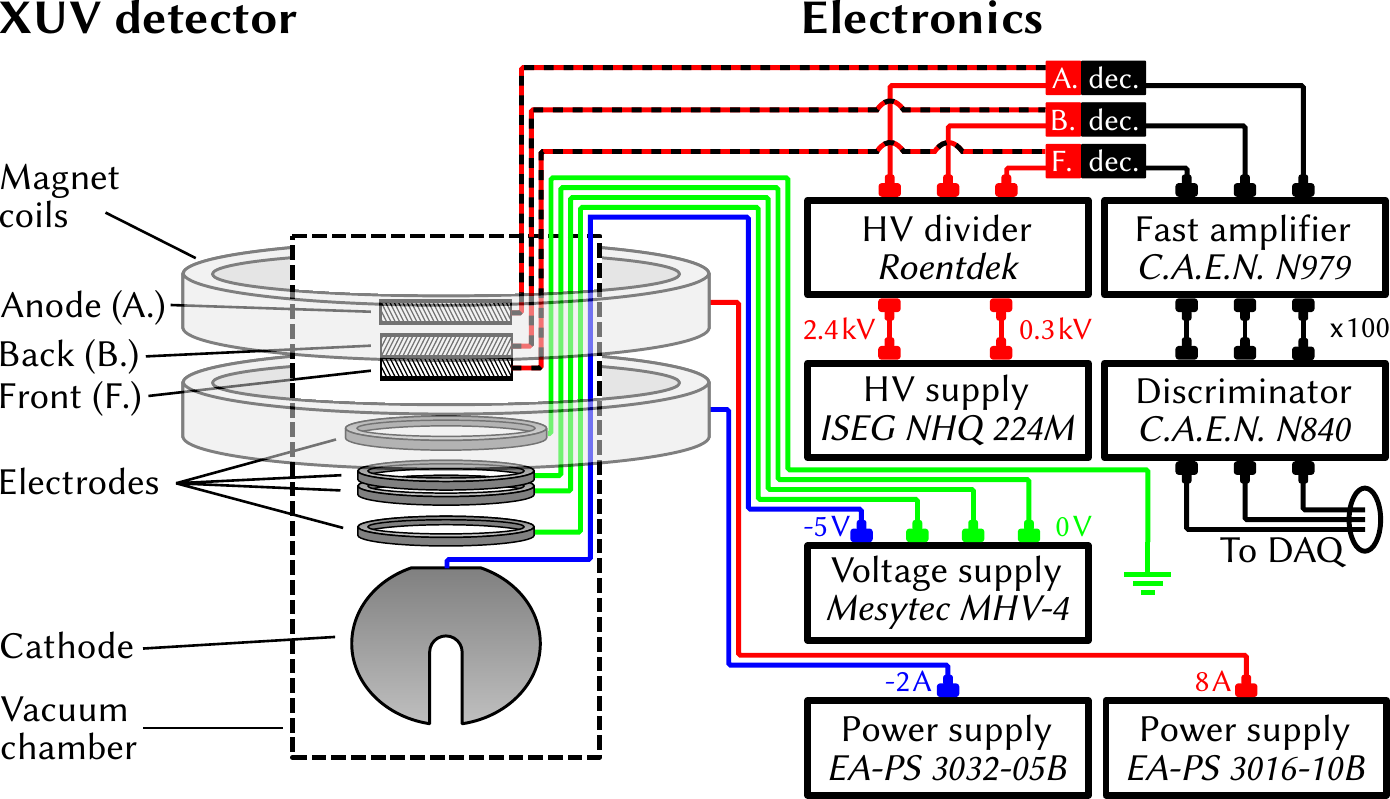}
\caption{Interconnections of the XUV detection system with power supplies and front-end electronics of the data acquisition system.}
\label{fig:xuv2}
\end{figure}

The measurements were performed with a beam of $^{12}\mathrm{C}^{3+}$-ions that was overlapped with an anti-collinear laser beam in one of the straight sections of the ESR to excite either the \p1 or the \p3 transition (see figure~\ref{fig:esr}).
For this purpose two laser systems were available: a pulsed laser system from HZDR/TU-Dresden~\cite{Sie16} and a continuous-wave laser system from TU-Darmstadt~\cite{Bec16}, both with a wavelength of $\lambda_{\rm lab} \approx 257\,\mathrm{nm}$. 
As the cw laser system provided a higher output power and a more narrow linewidth than the pulsed laser, it was used during most of the beamtime, including the measurements detailed in this paper.
According to the equation for the relativistic Doppler shift
\begin{equation}
  \lambda_0 = \lambda_{\rm lab} \cdot \gamma (1 + \beta\cos{\theta}) \, ,
  \label{eq:doppler}
\end{equation}
with the ion velocity $\beta~\approx~0.47$, the Lorentz-factor $\gamma = 1/\sqrt{1-\beta^2}$ and an angle between laser and ion beam of $\theta \approx 180^\circ$, the laser photons are blue-shifted to wavelengths around $\lambda_0 \approx 155 \,\mathrm{nm}$ in the rest frame of the ions. 
In order to obtain a well-determined absolute ion energy ($E_{\rm kin} = 122\;$MeV/u) and a small relative momentum spread ($dp/p \approx 1\cdot 10^{-5}$), the ion beam was cooled by the ESR electron cooler~\cite{Fra87}. Following the principle of electron cooling, the precise velocity of the ions is determined by the acceleration voltage of the electron beam inside the cooler. Scans across the two transitions of interest were realized by varying the electron cooler voltage, thus changing the velocity of the ions stored inside the ESR and thereby the Doppler-shifted laser wavelength in the rest frame of the ions. 
The dependence of the ion velocity on the effective electron cooler voltage $U_{\rm ecool}$ is given by 
\begin{equation}
  \beta_{\rm ion} = \sqrt{ 1 - \left( 1 - \frac{e \cdot U_{\rm ecool}}{m_e \cdot c^2}\right)^{-2}} \; ,
  \label{eq:beta}
\end{equation}
where $e$ and $m_e$ are the elementary charge and electron mass, respectively.
The acceleration voltage of the electron cooler was provided by a Heinzinger HNC~320.000-10 power supply and set on a terminal in the main control room of the storage ring.
Monitoring of the actual voltage present at the cooler was achieved using a Julie Research JRL~HVA-100 high-precision voltage divider connected to a 6.5 digit  Keysight~34465A digital voltmeter (DVM). \\
Fluorescence photons, which are emitted when the resonance condition for one of the two transitions is met, were detected by the new XUV detection system located in the region of overlap between the laser and the ion beam. Due to the Lorentz boost, the fluorescence photons are mainly emitted in the forward direction with the photon wavelengths being Doppler shifted down to $\lambda_{\rm fluo} (0^\circ)\approx 93~\,\mathrm{nm}$ (see inlay in figure~\ref{fig:esr}).
%
%
\paragraph{CW Laser system}
The measurements discussed in the following sections were performed using the continuous-wave (CW) laser system originally developed in the scope of a PhD thesis~\cite{Bec16,Bec15} and further optimized as reported in~\cite{Kie19, Klam17}. The setup is divided into five main components (see figure~\ref{fig:laser}).
A diode laser with an external resonator (external cavity diode laser, ECDL) serves as master oscillator operating at 1028~nm. The ECDL setup of Littrow design features a blazed grating that can be tilted via a piezoelectric actuator so that the output frequency of the laser system is tunable. 
The laser light from the ECDL is then amplified in an Ytterbium-doped fiber-amplifier system. 
Subsequently, the amplified light is frequency-doubled twice. The first frequency doubling is achieved by guiding the light in a single pass arrangement through a magnesium-doped periodically-poled lithium-niobate (MgO:PPLN) crystal.
From there, the emerging green light of approximately 514~nm is guided into the UV cavity, where it is again frequency-doubled with a $\beta$-barium borate (BBO) crystal.
The resulting beam with $\approx 257$~nm wavelength is then transported to the overlap region of laser and ion beams. 
In order to measure the output power during the beam-time, a fraction of the laser light was diverted by a beam splitter at the output of the UV cavity. As seen in figure~\ref{fig:laser}, it was continously monitored by a Gentec-EO power detector model XLP12-1S-H2-D0 and the data was stored on a local laptop. 
The laser system provides 4~W of infrared power before the first frequency doubling that results in about 1~W of power at 514~nm. After the BBO crystal an average output power of 20~mW in the UV was obtained.
To stabilize the laser frequency, a small fraction of the beam from the ECDL was coupled into a single-mode fiber and guided to a wavelength meter (WLM) from HighFinesse, type WS7-60. In a feedback loop deviations from the desired laser frequency measured by the device were then used to actuate the blazed grating, counteracting the fluctuation.
In this way we obtained a frequency stability of the infrared laser light of about 5~MHz at 291~THz. The beam diameter directly after the laser system was 3~mm  and about 10~mm in the overlap region between laser beam and ion beam.
%
%
\paragraph{XUV detection system}
For an effective detection of forward emitted XUV photons in the ESR a novel detection system was developed based on the idea of detecting secondary electrons produced by the incoming photons from a suitable cathode material~\cite{Vol16,Ege16}. The detector consists of a movable cathode plate that can be driven into the beam line via a pressured air motor. In order not to interfere with the ion beam, the plate has a central 30~mm wide slit. XUV photons hitting the cathode will produce mostly low energetic (typically a few eV) secondary electrons that will be guided electromagnetically onto a microchannel plate (RoentDek DET40 MCP) in Chevron configuration, which is mounted inside the vacuum. A similar detection system for optical photons, making use of a movable parabolic copper mirror and a photomultiplier outside the vacuum~\cite{Han13}, was already successfully applied in a measurement of the hyperfine structure splitting in H-like and Li-like bismuth~\cite{Loc14, Ull17a}.\\
Figure~\ref{fig:xuv1}, left panel, shows a schematic view of the setup as it was implemented for simulations using the software package SIMION~8.1~\cite{Simion}. 
The cathode is coated with a 300\;nm thick layer of caesium iodide (see figure~\ref{fig:xuv1}, right panel), vacuum evaporated onto the cathode surface in the target laboratory at GSI, to enhance the secondary electron yield at the wavelengths of interest. 
The yield of photocathodes with a 300~nm coating of CsI was measured by Henke et~al.~\cite{Hen81} and was found to vary between 0.3 and 2.5 electrons per incident photon over a photon energy range from 100~eV to 10~keV. For the UV region data exist for CsI coated MCPs that provide quantum efficiencies between 20\% and 30\% for wavelengths between 20~nm and 150~nm~\cite{Mar82}. Although the quantum efficiency drops sharply for longer wavelengths, we were able to successfully test the detector using UV light at a wavelength of 265~nm~\cite{Vol16}.\\
A magnetic field, produced by solenoid coils mounted outside the vacuum, is used to guide the electrons to the MCP. 
The simulations resulted in an optimum configuration with one solenoid coil providing the main guiding field whereas a second solenoid coil closer to the cathode is operated with a significantly smaller current in opposite direction, widening the magnetic field lines to cover the complete cathode plate (see blue field lines in figure~\ref{fig:xuv1}, left panel).
Ring electrodes (labeled (2) in the figure) were meant to provide additional electrostatic guiding of the electrons but were found to have only little effect on the collection efficiency of the system and were grounded during the experiment. The collection efficiency, i.e. the percentage of electrons that are successfully guided from the cathode plate onto the active diameter of the MCP, was found to reach 80\% with optimal settings of magnetic fields and cathode-plate voltage~\cite{Ege16}.\\
Figure~\ref{fig:xuv2} shows the interconnections of the detection system with the voltage supplies of the electrodes, the current supplies for the two solenoid coils and the front-end electronics for readout of the MCP.
The high voltage for the channelplates is generated using an ISEG NHQ 224M HV supply. Charge pulses amplified in the MCP front and back plates and detected in the MCP anode are capacitively decoupled from the high voltage lines and amplified using a CAEN N979 fast amplifier. After converting the analog pulses to logical NIM signals using a CAEN N840 Constant Fraction Discriminator the latter are recorded using a GSI VUPROM (VME Universal Processing Module) module~\cite{Hof06} serving both as a multi-hit TDC and a scaler module.
\section{Calibration and systematic uncertainties}
\label{sec:calibration}
A careful calibration of the different systems utilized in the setup must be carried out to minimize the systematic uncertainties in the deduced transition wavelengths.
A short overview of the most important calibration procedures is given in the following.
\paragraph{HV divider scale factor}
The uncertainty of the scale factor of the JRL HVA-100 high voltage divider used to monitor the electron cooler voltage is one of the main systematic uncertainties in the transition wavelength analysis presented in section~\ref{sec:analysis}.
For the HVA-100 divider, we have three sets of calibration data obtained with different reference dividers in 2013~\cite{Bau13}, 2014~\cite{Ull14} and 2017~\cite{Kon18} at voltages similar to the -67~kV applied during the beam-time.
Applying a linear fit to the scale factor as function of time, a scale factor for the measurement period in 2016 is extracted:
\begin{equation}
 M_{HVA-100}(-67\,{\rm kV},\, 2016) = 9999.698(155) \; .
\end{equation}
The relatively large uncertainty of 16~ppm is mainly due to missing temperature monitoring of the divider during the 2016 beam-time (the divider exhibits a dependence of the scale factor on the environmental temperature of $-2.75(25)\;$ppm/K~\cite{Kon18}).
\paragraph{DVM gain calibration}
A calibration of the Keysight 34465A DVM was performed in advance of the beam-time to determine a possible offset and gain factor of the device. 
The offset was found to be $0.000000(20)\;$V. The gain is determined utilizing a Fluke~732B reference source, which provides an output voltage of $10.000076(50)$~V, and was measured to be 
\begin{equation}
  G_{\rm DVM} = \frac{10.000076(50)\,{\rm V}}{U_{\rm DVM} - 0.000000(20)\,{\rm V}} = 1.0000006(131) \, {\rm V} \, ,
\end{equation}
where $U_{\rm DVM}$ is the voltage value read from the DVM.
\paragraph{Voltage dependence on cooler current}
The cooler voltage, as measured by the DVM, exhibits a small dependence on the cooler current. This dependence was measured at the two central cooler voltages used in the beamtime over a range from 0~mA to 250~mA cooler current.
Up to a cooler current of $122\;$mA we observe a linear drop in the measured DVM voltages by $-0.6\;\mu\,$V/mA after which it saturates at a value of $\approx -73\mu\,$V. This correction is used when reconstructing missing DVM measurements at different cooler currents.
\paragraph{Set voltage calibration}
Owing to problems with the readout software of the Keysight voltmeter, the measured electron cooler voltages were not written to storage for larger parts of the beam-time. This especially concerns data taken for the \p1 transition. 
For these parts of the data, we have to reconstruct the DVM measurements from the cooler voltage set values $U_{\rm set}$. To do so, we use periods of the beam-time where both DVM data and control room set voltages are available to develop a suitable conversion formula.
As these data were collected at different electron cooler currents, we calculate the corresponding DVM values at zero cooler current $U_{\rm DVM, \, 0\,mA}$ by applying the correction explained in the previous paragraph. \\
Besides a small offset of the set voltage and a scale factor, the formula has to take into account the finite 18~bit resolution with which the set voltages were transfered to the Heinzinger power supply of the electron cooler. The maximum output voltage of the supply is 320~kV so the least significant bit of the digital set voltage corresponds to $\Delta U = 320\,{\rm kV}/2^{18} = 1.220703\,{\rm V}$. With this the conversion formula was found to be
\begin{equation}
 U_{\rm DVM,\,0mA,\,recon.} = -\verb|floor|\left(\frac{U_{\rm set} + 0.38\,{\rm V}}{\Delta U}\right) \cdot \frac{\Delta U}{10005.6699} \; ,
\label{eq:uset}
\end{equation}
where the $\verb|floor|$ function returns the integer part of its operand. 
The relative deviation of measured DVM voltages $U_{\rm DVM, \, 0\,mA}$ and reconstructed values $U_{\rm DVM,\,0mA,\,recon.}$ stays below 10~ppm for all available datapoints.
\paragraph{Electron cooler workfunction}
The work function $W$ of a material has a direct impact on the electrostatic potential $\phi$ produced in the vacuum when a voltage $U$ is applied to the material: $\phi = U - W/e$. 
Electrons in the ESR electron cooler are generated in the dispenser cathode of an electron gun. It consists of a tungsten filament coated with barium, having a nominal work function of $W_{\rm W, Ba} \approx 1.66$~eV~\cite{Dem05}. The beampipe of the electron cooler, where the ion beam is superimposed with the electron beam, consists of stainless steel with a nominal work function of $W_{\rm steel} \approx 4.5$~eV~\cite{Lid96}.
Work functions are typically determined under ideal conditions with ultrapure surfaces. Work functions encountered under standard conditions can, therefore, deviate by up to $\approx 1$~eV from the literature values~\cite{Beh16}.
The acceleration voltage felt by the electrons inside the cooler is determined by the difference of the electrode potentials of cathode and beampipe $\Delta \phi$ given by
\begin{equation}
 \begin{aligned}
  \Delta\phi &= \phi_{\rm cath}-\phi_{\rm pipe}
   = \left(U_{\rm cath}-\frac{W_{\rm W,Ba}}{e}\right)-\left(U_{\rm pipe}-\frac{W_{\rm steel}}{e}\right) 
   = (U_{\rm cath}-U_{\rm pipe})-\left(\frac{W_{\rm W,Ba}}{e}-\frac{W_{\rm steel}}{e}\right) \\
  &= U_{\rm HNC} + 2.84(142)\,{\rm V} \; ,
 \end{aligned}
\end{equation}
where the difference between the cathode voltage $U_{\rm cath}$ and the beampipe voltage $U_{\rm pipe}$ is the voltage $U_{\rm HNC}$ applied to the electron cooler using the Heinzinger power supply. \\
In conclusion, the potential of the electron cooler is shifted by $+2.84(142)$~V, effectively lowering the electron energies. Considering an electron cooler voltage of approximately -67 kV, the uncertainty of the work function difference on this voltage amounts to 21~ppm, representing one of the major 
contributions to the error budget.
\paragraph{Space charge corrections}
Electron and ion space charges, present in the overlap region of both beams inside the electron cooler, modify the electric potential that accelerates the electron beam and therefore impact the velocity of the cooled ions. Whereas the effect of the electron space charges can be measured (see section~\ref{sec:current_corr}), the potential change due to the ion current has been estimated from the beam parameters and the drift tube geometry to be $\Delta \phi_{\rm ion} \approx -0.5$~V/mA~\cite{Win20}. 
To assess the effect of the ion space charge, the analysis has been performed with and without the correction resulting in a $\approx 0.3$~ppm change in the extracted wavelength, which was then used as a conservative estimate of the related systematic uncertainty.
\paragraph{Laser wavelength}
Two different wavelength meters were used to determine and stabilize the frequency of the CW-laser. 
1) A HighFinesse WS6-200 wavelength meter, calibrated to a HeNe-laser, was used to measure the absolute frequency of the ECDL driving the system before and after the experiment. 
2) A HighFinesse WS7-60 wavelength meter was utilized for a continuous stabilization of the laser frequency. 
Both wavelength meters operate in a range of 330~nm to 1180~nm and mainly differ in terms of accuracy. The WS6-200 has a 1 sigma uncertainty of 67~MHz and the WS7-60 has a 1 sigma uncertainty of 20~MHz.
The measurement of the ECDL frequency with the calibrated WS6-200 amounted to $f_{\rm laser} = 291.43340(7)$~THz. The frequency is quadrupled in a two-step process and thus, the resulting laser wavelength is given by $\lambda_{\rm laser} = c/(4\cdot f_{\rm laser}) = 257.170642(62)$~nm .
\paragraph{Angle between laser and ion beam}
An additional uncertainty arises from the unknown angle $\theta$ between laser and ion beam. The effective laser wavelength in the ions rest frame directly depends on this angle according to equation~\ref{eq:doppler}. 
The position of the ion beam in the vacuum beam-pipe of the ESR can be determined using two sets (horizontal and vertical) of position-calibrated “scrapers” inside the vacuum, which are a length $l=6.33\;$m apart. These scrapers are simple metal plates that can be moved with sub-millimetre precision, blocking part of the ion beam. By moving them from the outside towards the center of the beam-pipe, the radius and the central position of the beam can be determined.
The same scrapers are also used to measure the position of the laser beam. The centre points of laser and ion beam - at the positions of the scrapers - were at maximum a distance $d=2\;$mm apart. The misalignment angle between laser and ion beam is therefore estimated to be $\Delta\theta_{\rm max} = \arctan(2\cdot d/l) = 0.036^\circ$ causing a corresponding uncertainty in the transition wavelengths of 0.000027\;nm for both transitions. 
\section{Resonance analysis}
\label{sec:analysis}
\begin{figure}[h]
\centering
\includegraphics[width=\textwidth]{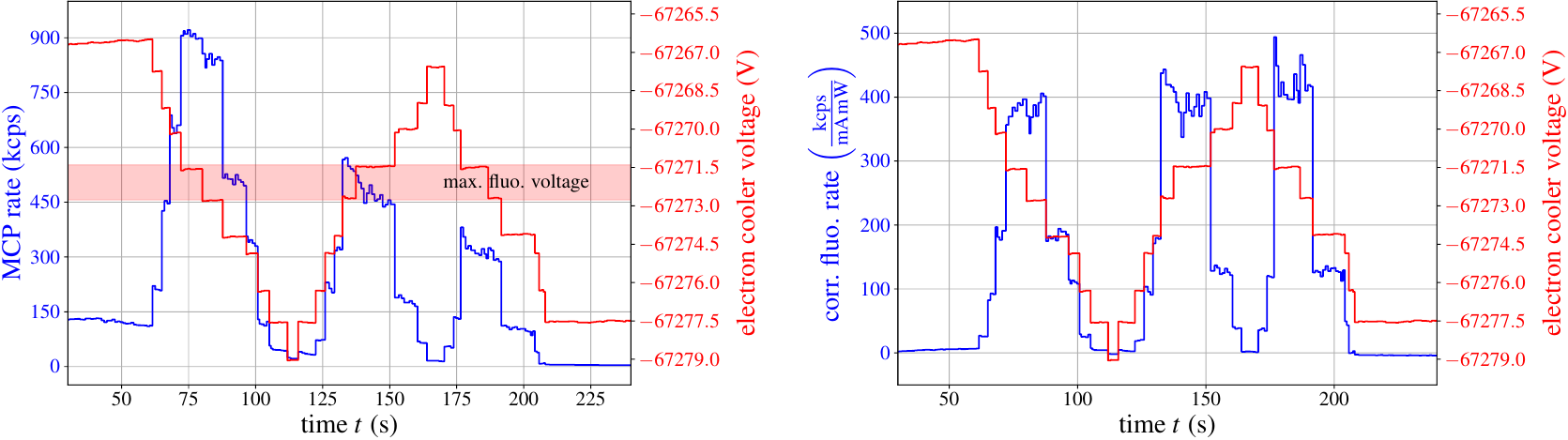}
\caption{Left: raw detector rates (blue curve) for a scan of the electron cooler voltage (red curve) over the resonance region of the \p3 transition. Right: fluorescence rates of the same voltage scan after background subtraction, normalization to ion current and laser power and MCP gain correction.}
\label{fig:scan}
\end{figure}

\begin{figure}[t]
\centering
\includegraphics[width=\textwidth]{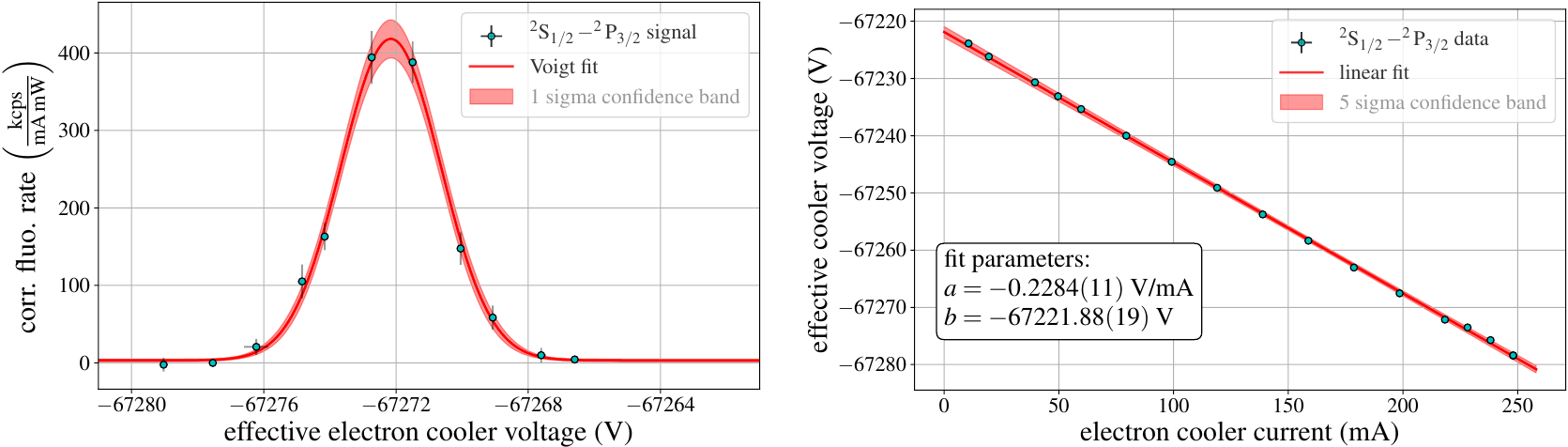}
\caption{Left: Observed resonance curve for the \p3 transition at a cooler current of 218~mA. Right: Measured resonance voltages as a function of the cooler current for the \p3 transition.}
\label{fig:final}
\end{figure}

The following sections provide an overview of the experimental procedure during the beamtime and the extraction of the resonance wavelengths for the \p1 and \p3 transitions.
%
\paragraph{Experimental background}
There are two major sources of experimental background for the fluorescence detection system. One is laser stray light creating photoelectrons from the cathode plate that are subsequently guided to the MCP. This leads to a constant background whose magnitude depends on the position of the cathode plate inside the beam line.
The second source of background is caused by the ion beam ionizing or exciting residual gas molecules inside the beam pipe, which can lead to electrons or photo-electrons, respectively, finally hitting the MCP. The magnitude of this background component is proportional to the ion beam current.
During initial tests of the detector in this beam-time, it was found that the detector could not be used with the cathode plate positioned around the ion beam in the center of the beam pipe due to excessive background rates that saturated the MCP. Therefore, the cathode was positioned at a distance of $\approx 10\;$cm from the ion beam.
The resulting background contributions were then found to be $b_{\rm ion} = 146\,$kcps/mA from the ion beam and $b_{\rm laser} = 0.11\,$kcps/mW from the laser system~\cite{Win20}.
%
\paragraph{Voltage scans}
\label{sec:scans}
To determine the exact wavelengths of the \p1 and \p3 transitions, scans of the electron cooler voltage were performed thus changing the velocity of the stored ions inside the ESR and thereby the Doppler-shifted laser wavelength in the rest frame of the ions. Fluorescence photons that were created when the resonance condition for either of the two transitions was met, were detected and recorded by the data acquisition system. Figure~\ref{fig:scan} shows on the left panel the raw MCP rates $R_{\rm MCP}$ (blue curve) recorded while scanning the cooler voltage (red curve) in 1\,V steps three times over the \p3 resonance. 
The maximum rates, corresponding to the resonance condition being fulfilled, occur for a cooler voltage of approximately $-67272$~V. 
The observed decrease in the peak amplitudes is caused by the loss of ions in the ring. The beam current was continously monitored using a DC current transformer in the beam diagnostic system. During the measurements we observed a mean lifetime of the beam current of about 62 seconds.\\
To extract the exact resonance voltage, the experimental background is subtracted from the data which are subsequently normalized to the measured ion current $I_{\rm ion}$ and laser power $P_{\rm laser}$. Finally a gain correction $G_{\rm MCP}$ compensating for saturation effects of the MCP at high event rates is applied 
to obtain the fluorescence rate $R_{\rm fluo}$
\begin{equation}
   R_{\rm fluo} = \frac{R_{\rm MCP} - b_{\rm ion} \, I_{\rm ion} - b_{\rm laser}\, P_{\rm laser}}{I_{\rm ion} \, P_{\rm laser}} 
                  \frac{1}{G_{\rm MCP}}\; .
   \label{eq:rate_corr}
\end{equation}
The gain correction can, following methods by Giudicotti~\cite{Giu94} and Gershman~\cite{Ger18}, be written as 
\begin{equation}
 G_{\rm MCP} = \exp\left( - \frac{R_{\rm MCP}}{\epsilon \cdot R_{50\%}}\right) \; ,
\end{equation}
where the MCP efficiency is stated by the manufacturer to be $\epsilon \approx 0.6$ for electrons with energies between 100~eV and 1000~eV~\cite{Roe20} (the electrons from the cathode are accelerated by a positive voltage of 300~V applied to the front-plate of the MCP). $R_{50\%}$ corresponds to the rate for which the MCP gain drops to 50\% of the unsaturated value and has been determined to be $R_{50\%} = 1603(32)\,$kcps~\cite{Win20}. 
The effect of the gain correction on the reconstructed fluorescence rate $R_{\rm fluo}$ reaches a factor 2 at an MCP rate of $R_{\rm MCP}\approx 660$~kcps.
The corrected fluorescence rates, obtained by applying equation~\ref{eq:rate_corr} to the raw rates, are displayed in figure~\ref{fig:scan}, right panel. The similar height obtained for all three peaks provides confidence in the normalizing procedure.\\
Accumulating all events from a given experimental run that belong to the same voltage step and plotting the corrected fluorescence rates against the effective electron cooler voltage (containing all corrections discussed in section~\ref{sec:calibration}), one obtains a resonance curve as displayed in figure~\ref{fig:final}, left panel. By fitting a Voigt profile to the data points, the precise voltage of the peak center is determined, together with its associated uncertainty.
%
%
%
\paragraph{Electron space charge correction}
\label{sec:current_corr}
To account for the electron space charge inside the ESR cooler, which modifies the acceleration potential felt by the electrons and therefore the ion velocity, the transition measurements were performed for a series of electron cooler currents, ranging from 10~mA to 250~mA. Figure~\ref{fig:final} displays the observed resonance voltages for the \p3 transition (right panel) as a function of the electron cooler current.
Performing linear fits to the two datasets allows to extrapolate the measurements to zero cooler current, thereby eliminating the influence of the electron space charge effect.
The resulting resonance voltages for the two transitions are $U_0(^2{\rm P}_{1/2}) = -66774.13(11)\,{\rm V}$ and $U_0(^2{\rm P}_{3/2}) = -67221.88(19)\,{\rm V}$.
%
%
\paragraph{Transition wavelengths}
From the resonance voltages of the two transitions at zero cooler current given above, the transition wavelength in the ions rest frame can be calculated according to equations~\ref{eq:doppler} and~\ref{eq:beta}, yielding
$\lambda_0(^2{\rm P}_{1/2}) = 155.0779(12)_{sys}(1)_{stat}\,{\rm nm}$ and
$\lambda_0(^2{\rm P}_{3/2}) = 154.8211(12)_{sys}(2)_{stat}\,{\rm nm}$, 
where the statistical uncertainties result from the fit uncertainties of the resonance voltages and the systematic uncertainties result from the different contributions discussed in section~\ref{sec:calibration}
\begin{table}[h]
 \caption{Contributions of the individual systematic uncertainties to the overall error budget. A variable parameter is labeled 'var.'. If the uncertainty of a parameter is labeled 'max.', the corresponding contribution to the error budget of the wavelength has been estimated by omitting the correction from the analysis.}
 \label{tab:systematic}
  \begin{tabular}{llllcc}
   \hline
    Parameter              & value     & uncertainty & unit     & \multicolumn{2}{c}{error contribution [nm]} \\
    \rule{0pt}{2ex}        &           &             &          & \p1      & \p3  \\
    \hline
    work function          & 2.84      & 1.42        & eV       &  0.00082 & 0.00081 \\
    scale factor           & 9999.698  & 0.155       &          &  0.00060 & 0.00060 \\  
    DVM gain               & 1.0000006 & 0.0000131   &          &  0.00050 & 0.00050 \\
    set voltage cal.       & var.      & 0.000067    & V        &  0.00038 & -       \\    
    e-cooler current       & var.      & 2           & mA       &  0.00026 & 0.00026 \\
    DVM offset             & 0.0       & 0.00002     & V        &  0.00012 & 0.00011 \\
    ion space charge       & var.      & max.        & V        &  0.00005 & 0.00004 \\
    laser frequency        & 291.43340 & 0.00007     & THz      &  0.00004 & 0.00004 \\
    angle laser - ion beam & 0.0       & 0.036       & $^\circ$ &  0.00003 & 0.00003 \\
    MCP gain               & var.      & max.        &          &  0.00001 & 0.00003 \\
    \hline
    \hline
    root summed squares                                      &&&&  0.00123 & 0.00116 \\
  \end{tabular}
\end{table}
and given in detail in table~\ref{tab:systematic} for the two transitions. \\
Table~\ref{tab:comparison} provides a comparison of the results of this work to previous measurements and theoretical predictions.
\begin{table}[h]
 \caption{Overview of measured and calculated wavelengths for the \p1 and \p3 transitions obtained in this work and available from the literature.}
 \label{tab:comparison}
  \begin{tabular}{lllll}
    \hline
                                  & year & method                  & $\lambda(^2{\rm P}_{1/2})$ [nm]                                        & $\lambda(^2{\rm P}_{3/2})$ [nm] \\ 
    \hline
    Experiment: \\
    {\bf this work}               & {\bf 2020} & {\bf laser spectroscopy} & $\mathbf{155.0779(12)_{sys}(1)_{stat}}$ & $\mathbf{154.8211(12)_{sys}(2)_{stat}}$ \\
    Schramm et al.~\cite{Sch05}   & 2005       & laser spectroscopy       & $155.0705(39)_{\rm sys}(3)_{\rm stat}$                     & $154.8127(39)_{\rm sys}(2)_{\rm stat}$ \\
    Griesmann et al.~\cite{Gri00} & 2000       & interferometry           & 155.0781(2)                                                & 154.8204(1) \\
    Bockasten et al.~\cite{Boc63} & 1963       & plasma spectroscopy      & 155.0774(10)                                               & 154.8202(10) \\
    \hline
    Theory: \\
    Yerokhin et al.~\cite{Yer17}  & 2017       & RCI$^a$           & 155.080(29)   & 154.827(36) \\
    Borschevsky~\cite{Bor14}      & 2014       & FSCC$^b$          & 155.075       & 154.820 \\
    Tupitsyn et al.~\cite{Sch05}  & 2003       & n/a               & 155.0739(26)  & 154.8173(53) \\
    Johnson et al.~\cite{Joh96}   & 1996       & RMBPT$^c$         & 155.078       & 154.819 \\
    Kim et al.~\cite{Kim91}       & 1991       & RMBPT+MCDHF$^d$   & 155.060       & 154.804 \\
    \hline
    \multicolumn{5}{l}{\small $^a$ Relativistic Configuration Interaction; $^b$ Fock Space Coupled Cluster; $^c$ Relativistic Many Body Perturbation Theory;} \\
    \multicolumn{5}{l}{\small $^d$ MultiConfiguration Dirac-Hartree-Fock} \\
  \end{tabular}%
\end{table}
The wavelengths determined for the two transitions are in good agreement with the experimental results obtained from interferometry~\cite{Gri00} and plasma spectroscopy~\cite{Boc63} and, within the uncertainties, also with the given theoretical predictions. It was possible to improve the precision with respect to the previous ESR measurement~\cite{Sch05} by about a factor of three.\\[2mm]
%
A further improvement in accuracy could be achieved in future measurements by tackling the three most important systematic uncertainties:
1) Measuring the work function difference between cooler electron gun and drift electrodes, e.g. by performing laser spectrocopy on ions with
a well-known transition. 
2) Using a freshly calibrated high-voltage divider for the measurement of the cooler voltage and monitoring or stabilizing its temperature during the experiment. 
3) Using a 8.5 digit DVM to measure the cooler voltage. 
While the achievable accuracy of a future work function measurement is difficult to estimate, the accuracy of the scale factor of the divider used in this measurement could be improved to the <5~ppm level assuming a temperature stability of $1^\circ$C, as was shown in the 2013 calibration of the device~\cite{Bau13}.
An 8.5 digit precision DVM together with a 10~V reference source (like e.g. the Fluke 732C) would allow for a 1~ppm precise readout of the divider output voltage. \\[2mm]
On the experimental side, the new cw UV-laser system provided a stable and reliable beam to the experiment and the novel XUV detection system proved to have a high detection efficiency for Lorentz-boosted fluorescence photons. To lower the background sensitivity of the XUV detector, dedicated shielding electrodes and a stepper-motor driven motion of the cathode are being implemented.
{}
\section*{Acknowledgements}
This work was supported by the German Federal Ministry of Education and Research (BMBF)
under grant numbers 05P15PMFAA, 05P18RDFAA, 05P09RDFA3, 05P12RDRB2, 05P15RDFA1 and 05P16ODFA1.
D.W. and J.U. acknowledge support from HGS-HiRe.
We appreciate the technical support during the beam-time by the GSI accelerator department.
\section*{Author contributions statement}
%
%
D.F.A.W. conceived and organised the experiment.
D.K., S.K., and T.W. developed the TU Darmstadt CW laser system.
M.S. and M.L. developed the TU Dresden pulsed laser system.
D.W., V.H., C.E., H.-W.O., J.V. and C.W. developed the XUV detector.
L.E., R.S. and J.U. prepared the data acquisition system. 
W.N. and J.U. provided the HV divider.
D.W., V.H., M.B., A.B., C.E., D.K., S.K., T.K., R.S., J.U., HB.W., ZQ.H. , L.E. and D.F.A.W. worked on experimental shifts.
D.W. performed the analysis of the data.
M.B., X.M., W.N., T.S., T.W. and C.W. are group leaders.
V.H., D.W. and D.F.A.W. prepared the manuscript.
All authors reviewed the manuscript.
\section*{Competing interests}
The authors declare no competing interests.
\end{document}